\documentclass[aps,amsmath,showpacs,amsfonts,10pt]{revtex4}
\usepackage{epsfig,graphicx}
\usepackage[english]{babel}
\usepackage{amsfonts}
\usepackage{amsmath}
\usepackage{latexsym}
\usepackage{graphics,bm}
\usepackage{natbib}
\usepackage{dcolumn}
\usepackage{bm}
\usepackage{rotating}

\begin{document}

\title{Optomechanical coupling between two optical cavities: cooling of a micro-mirror and parametric normal mode splitting}

\author{Tarun Kumar$^{1}$, Aranya B.\ Bhattacherjee$^{2}$ and ManMohan$^{1}$}

\address{$^{1}$Department of Physics and Astrophysics, University of Delhi, Delhi-110007, India} \address{$^{2}$Department of Physics, ARSD College, University of Delhi (South Campus), New Delhi-110021, India}

\begin{abstract}
We propose a technique aimed at cooling a harmonically oscillating mirror mechanically coupled to another vibrating mirror to its quantum mechanical ground state. Our method involves optmechanical coupling between two optical cavities. We show that the cooling can be controlled by the mechanical coupling strength between the two movable mirrors, the phase difference between the mechanical modes of the two oscillating mirrors and the photon number in each cavity. We also show that both mechanical and optical cooling can be achieved by transferring energy from one cavity to the other.  We also analyze the occurrence of normal-mode splitting (NMS). We find that a hybridization of the two oscillating mirrors with the fluctuations of the two driving optical fields occurs and leads to a splitting of the mechanical and optical fluctuation spectra.
\end{abstract}


\maketitle

\section{Introduction}

In recent years mechanical and optical degrees of freedom have become entangled experimentally by underlying mechanism of radiation pressure forces. This field known as cavity optomechanics has played a vital role in the conceptual exploration of the boundaries between classical and quantum mechanical systems. The coupling of mechanical and optical degrees of freedom via radiation pressure has been a subject of early research in the context of laser cooling \citep{hansch,wineland, chu} and gravitational-wave detectors \citep{caves}. Recently there has been a great surge of interest in the application of radiation forces to manipulate the center-of-mass motion of mechanical oscillators covering a huge range of scales from macroscopic mirrors in the Laser Interferometer Gravitational Wave Observatory (LIGO) project \citep{corbitt1, corbitt2} to nano-mechanical cantilevers\citep{hohberger, gigan, arcizet, kleckner, favero, regal}, vibrating microtoroids\citep{carmon, schliesser}, membranes\citep{thompson} and Bose-Einstein condensates \citep{brennecke, murch, bhattacherjee,bhattacherjee2}. The central accomplishment of the field of cavity optomechanics is the investigation of radiation pressure forces which allow one to manipulate the motional state of micromechanical oscillators. In particular, it has become possible to substantially cool the thermal excitation of a single mechanical mode, down to a few tens of remaining phonons \citep{schliesser_2}. With these developments, micro- and nanomechanical resonators now represent an important model system with the prospect of demonstrating quantum effects on a macroscopic scale. Theoretical work has also proposed to use the radiation-pressure coupling for quantum non-demolition measurements of the light field \citep{braginsky}.

It has been shown that ground state cooling of micro-mechanical mirror is possible only in the resolved side band regime (RSB) where the mechanical resonance frequency exceeds the bandwidth of the driving resonator \citep{marquardt,braginsky}. The cooling of mechanical oscillators in the RSB regime at high driving power can entail the appearance of normal mode splitting (NMS) \citep{dobrindt}. Recently, it was shown that an optical parametric amplifier inside a cavity considerably improves the cooling of a micro-mechanical mirror by radiation pressure \citep{huang}. Recently, dynamics of a micro mirror was studied in the presence of a nonlinear kerr medium placed inside an optical cavity \citep{tarun}. It was demonstrated that due to the photon blockade mechanism, as the Kerr nonlinearity is increased, the NMS progressively decreases. The Kerr medium was found to be a new handle to efficiently control the micro-mirror dynamics and this suggests a possible quantum device \citep{tarun}.

In this work, we propose a technique aimed at cooling a harmonically oscillating mirror (mechanically coupled to another vibrating mirror) to its quantum mechanical ground state. Our method involves optmechanical coupling between two optical cavities. We show that the cooling can be controlled by the mechanical coupling strength between the two movable mirrors and the phase difference between the mechanical modes of the two oscillating mirrors. We also analyze the occurrence of normal-mode splitting (NMS). We find that a hybridization of the two oscillating mirrors with the fluctuations of the two driving optical fields occurs and leads to a splitting of the mechanical and optical fluctuation spectra. The continuous variable entanglement between two mechanical modes could be used to improve the detection of weak classical forces in optomechanical devices as atomic force microscopes or gravitational wave detectors. Optomechanically coupled mirrors has been investigated earlier \citep{mancini,ludwig}. A continuous variable entanglement between the two mirrors was maintained by the light bouncing between the mirrors and was found to be robust against thermal noise \citep{mancini}. Entanglement between two mechanical oscillators coupled to a nonequilibrium environment showed that there is an optimal dissipation strength for which the entanglement between two coupled oscillators is maximized \citep{ludwig}. A new cooling method which involves the two-sided irradiation of the vibrating mirror inside an optical cavity has been proposed recently \citep{bhattacharya}. This method provides a stiffer trap for cooling the mirror and has several advantages over conventional methods of optomechanical cooling.

\section{Theoretical Framework}

	\indent We consider two Fabry-Perot cavities connected with each other through their movable mirrors as shown in fig. 1. Here mirror M1 and mirror M4 are fixed and are partially transmitting whereas mirrors M2 and M3 are movable and totally reflecting. The two mirrors M2 and M3 can both oscillate under the effect of the radiation pressure. The motion of each mirror is the result of the excitation of many oscillation modes which can be either external or internal. The external modes corresponds to pendulum modes which leads to global displacements of the mirror while the internal modes corresponds to deformations of the mirror surface due to excitation of internal acoustic modes of the mirror surface. These various degrees of freedom have different resonance frequencies and experimentally it is possible to select the mechanical response of a single mode by using a bandpass filter in the detection circuit. Consequently, we will consider a single mechanical mode for each movable mirror, which will be therefore described as a simple harmonic oscillator. The system under consideration is in contact with thermal bath in equilibrium with thermal bath at temperature T. The movable mirrors are treated as quantum mechanical oscillators with masses m1 and m2, frequencies $\nu1$ and $\nu2$, and energy decay rates $\gamma_{m1}$ and $\gamma_{m2}$ respectively of the mirrors M1 and M4. The system is coherently driven by two laser fields ($a_{1,in}$ and $a_{2,in}$) with frequencies $\omega_{1}$ and $\omega_{2}$ as shown in fig. 1. It is well known that high Q-optical cavities can significantly isolate the system from its environment, thus strongly reducing decoherence and ensuring that the light field remains quantum mechanical for the duration of the experiment. We also assume that the induced cavity resonance frequency shift of each cavity is much smaller than the longitudinal spacing, so that we restrict the model to a single longitudinal mode for each cavity. Let $\epsilon_{1}$ and $\epsilon_{2}$ be the amplitudes of the two laser fields. As we know in a Fabry-perot cavity, when a photon collides with the surface of the movable mirror, it exerts radiation pressure on the mirror and the force that mirror will experience is proportional to the photon number inside the cavity. But in our system here, the force experienced by one of the movable mirror, say M2 not only depends on the number on photons of the corresponding cavity but also depends on the number of photons of the second cavity. This is because the two mirrors are coupled, therefor position of one mirror is influenced by the position of the other mirror. We also assume that $\omega_{1}$, $\omega_{2}<<$ $\pi$c/L (adiabatic limit); c is the speed of light in vacuum and L is the cavity length in the absence of the cavity field. (assuming same length for the two cavities). The Hamiltonian of the system can be written as

\begin{figure}[t]

\hspace{-0.0cm}
\includegraphics [scale=0.6]{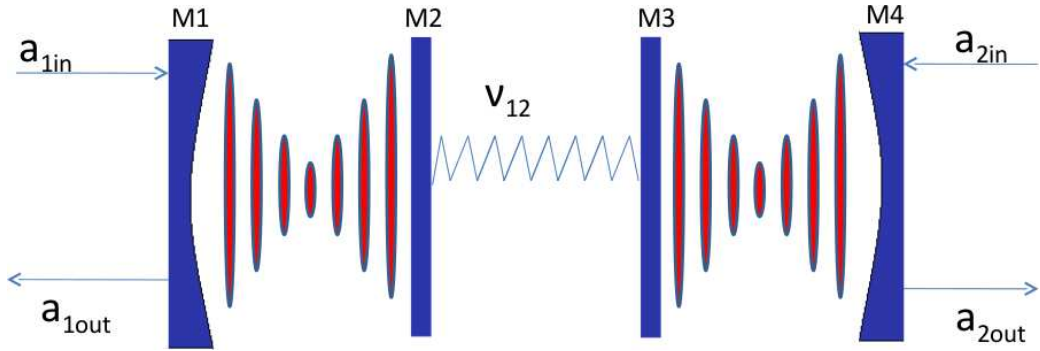}
\caption{Schematic description of the system under study.Two Fabry-Perot cavities are connected with each other through their movable mirrors. Here mirror M1 and mirror M4 are fixed and are partially transmitting whereas mirrors M2 and M3 are movable and totally reflecting. The system is coherently driven by two laser fields $a_{1,in}$ and $a_{2,in}$. $a_{1,out}$ and $a_{2,out}$ are the output fields.}
\label{1}

\end{figure}

\begin{eqnarray}\label{equation1}
H&=& \hbar \omega_{1}a_{1}^{\dagger}a_{1} + \hbar \omega_{2}a_{2}^{\dagger}a_{2} + \hbar \nu_{1}(b_{1}^{\dagger}b_{1}+1/2) + \hbar \nu_{2}(b_{2}^{\dagger}b_{2}+1/2) + \hbar \nu_{12}(b_{1}^{\dagger}e^{-i \theta_{1}} + b_{1}e^{i \theta_{1}})(b_{2}^{\dagger}e^{-i \theta_{2}} + b_{2}e^{i \theta_{2}}) \nonumber \\
&-& \hbar g_{1}a_{1}^{\dagger}a_{1} (b_{1}^{\dagger}e^{-i \theta_{1}} + b_{1}e^{ i \theta_{1}}) + i \hbar \epsilon_{1}(a_{1}^{\dagger}-a_{1}) - \hbar g_{2}a_{2}^{\dagger}a_{2} (b_{2}^{\dagger}e^{-i \theta_{2}} + b_{2}e^{i \theta_{2}}) +  i \hbar \epsilon_{2}(a_{2}^{\dagger}-a_{2})
\end{eqnarray}

Here $a_{1}$ ($a_{1}^{\dagger}$) and $a_{2}$ ($a_{2}^{\dagger}$) are the annihilation (creation) operators of the two cavity fields, $b_{1}$ ($b_{1}^{\dagger}$) and $b_{2}$ ($b_{2}^{\dagger}$) are the phonon annihilation (creation) operators of the two movable mirrors M2 and M3 respectively. The parameters $g_{1}$ and $g_{2}$ are the coupling parameters between the cavity fields and fixed mirrors M1 and M4 respectively, $\nu_{12}$ is the coupling frequency of the two movable mirrors and $\theta_{1}$ and $\theta_{2}$ are the phases of the two movable mirrors. The phases $\theta_{1}$ and $\theta_{2}$ can be thought of as arising from the complex mirror-photon coupling strengths $g_{i} (i=1,2)$.

The system we are considering here is intrinsically open as the cavity fields are damped by the photon leakage through the massive coupling mirrors. In the absence of the radiation pressure coupling, the cantilevers would undergo pure Brownian motion, driven by their contact with the thermal environment. The motion of the system can be described by the following Quantum Langevin equations-

\begin{equation}\label{equation1(a)}
\dot{q_{1}}=\nu_{1}p_{1}+Ap_{2}+Bq_{2}+2D_{2}|a_1|^{2}-\gamma_{m1}q_{1}
\end{equation}

\begin{equation}\label{equation1(b)}
\dot{q_{2}}=\nu_{2}p_{2}+Ap_{1}-Bq_{1}+2D_{4}|a_2|^{2}-\gamma_{m2}q_{2}
\end{equation}

\begin{equation}\label{equation1(c)}
\dot{p_{1}}=-\nu_{1}q_{1}-Aq_{2}+Bp_{2}+2D_{1}|a_1|^{2}-\gamma_{m1}p_{1}+\sqrt{2\gamma_{m1}}p_{in}^{1}
\end{equation}

\begin{equation}\label{equation1(d)}
\dot{p_{2}}=-\nu_{2}q_{2}-Aq_{1}-Bp_{1}+2D_{3}|a_2|^{2}-\gamma_{m2}p_{2}+\sqrt{2\gamma_{m2}}p_{in}^{2}
\end{equation}

\begin{equation}\label{equation1(e)}
\dot{a_{1}}=-i \omega_{1}a_{1}+ i D_{1}a_{1}q_{1}-i D_{2}a_{1}p_{1}-\kappa_{1}a_{1}+\sqrt{2\kappa_{1}}C_{in}^{1}+\epsilon_{1}
\end{equation}

\begin{equation}\label{equation1(f)}
\dot{a_{2}}=-i \omega_{2}a_{2}+ i D_{3}a_{2}q_{2}-i D_{4}a_{2}p_{2}-\kappa_{2}a_{2}+\sqrt{2\kappa_{2}}C_{in}^{2}+\epsilon_{2},
\end{equation}

where we have defined ($b_{i}$+$b_{i}^{\dagger}$)=$q_{i}$ and $i$($b_{i}^{\dagger}$-$b_{i}$)=$p_{i}$ ; $i=1,2$. Also $D_{1}$ = $g_{1}\cos(\theta_{1})$, $D_{2}$ = $g_{1}\sin(\theta_{1})$, $D_{3}$ = $g_{2}\cos(\theta_{2})$, $D_{4}$ = $g_{2}\sin(\theta_{2})$, $A$ = $\nu_{12}\cos(\theta_{2}-\theta{1})$, $B$ = $\nu_{12}\sin(\theta_{2}-\theta{1})$. $C_{in}^1$ and $C_{in}^2$ are input noise operators with zero mean value and obeys following commutation relation
$<\delta C_{in}^{i}\delta C_{in}^{j\dagger}>$ = $\delta_{ij}(t-t^{'})$, $<\delta C_{in}^{i}\delta C_{in}^{j}>$, $<\delta C_{in}^{i\dagger}\delta C_{in}^{j\dagger}> = 0$. Also $p_{in}^i$ = $ i(\xi^{i\dagger}-\xi^{i})$, $\xi$ is the Brownian noise operator, arising due to the thermal bath. Brownian noise operator has zero mean value and obeys following correlation at temperature $T$:  $<\xi^{i}(t)\xi^{j\dagger}(t^{'})> = 2\gamma_{mi}(1+2n_{T})\delta_{ij}(t-t^{'})$ and  $<\xi^{i}(t)\xi^{j}(t^{'})> = <\xi^{i\dagger}(t)\xi^{j\dagger}(t^{'})> = <\xi^{i\dagger}(t)\xi^{j}(t^{'})>=0$, $n_T = coth(\frac{\hbar\omega}{2k_BT})$, where $k_B$ is the Boltzmann constant and $T$ is the temperature of the thermal bath.

\section{Small Fluctuations Dynamics: Normal Mode Splitting and Cooling of a micro mirror}

Here we show that the coupling of the two mechanical oscillators and the two cavity field fluctuations leads to the splitting of the normal mode into two modes (Normal Mode Splitting(NMS)) for each cavity depending on the system parameters. The optomechanical NMS however involves driving four parametrically coupled nondegenerate modes out of equilibrium. The NMS does not appear in the steady state spectra but rather manifests itself in the fluctuation spectra of the mirror displacement. In order to study the dynamics of the coupled mirror, we need to find out the fluctuations in the mirror's position. As is clear from the equations 2-7 that the problem involved here is non-linear. We assume that this non-linearity is small. Therefor we study the dynamics of fluctuations around the steady state of the system.  We write each canonical operator of the system as a sum of its steady state mean value and a small fluctuation with zero mean value, $q_{1}\rightarrow q_{1s}+\delta q_{1}$,  $q_{2}\rightarrow q_{2s}+\delta q_{2}$, $p_{1}\rightarrow p_{1s}+\delta p_{2}$, $p_{2}\rightarrow p_{2s}+\delta p_{2}$, $a_{1}\rightarrow a_{1s}+\delta a_{1}$, $a_{2}\rightarrow a_{2s}+\delta a_{2}$. The steady state values are obtained by putting the left hand side of Eqns.(2)-(7) to zero. In order to achieve ground state cooling, we will always take $\gamma_{mi}$ $<<$ $\kappa_{i} $, $g_{i}$ $<$ $\nu_{i}$ and $\nu_{i}$ $>$ $\kappa_{i}$ (with $i=1,2$) The last condition is the resolved side band regime necessary for ground state cooling. Note that these conditions necessary for cooling also implies that the system is stable.  Linearizing equation 2 to 7 to obtain following Heisenberg - Langevin equations for the fluctuation operators :

\begin{equation}
\dot{\delta q_{1}}= \nu_{1}\delta p_{1} + A\delta p_{2} + B\delta q_{2} + 2D_{2}a_{1s}\delta a_{1}^{\dagger} + 2D_{2}a_{1s}^{*}\delta a_{1}
\end{equation}

\begin{equation}
\dot{\delta q_{2}}= \nu_{2}\delta p_{1} + A\delta p_{1} - B\delta q_{1} + 2D_{4}a_{2s}\delta a_{2}^{\dagger} + 2D_{4}a_{2s}^{*}\delta a_{2}
\end{equation}

\begin{equation}
\dot{\delta p_{1}}= -\nu_{1}\delta q_{1} - A\delta q_{2} + B\delta p_{2} + 2D_{1}a_{1s}\delta a_{1}^{\dagger} + 2D_{1}a_{1s}^{*}\delta a_{1}+ \sqrt{2\gamma_{m2}}\delta p_{in}^1-\gamma_{m1} \delta p_{1}
\end{equation}

\begin{equation}
\dot{\delta p_{2}}= -\nu_{2}\delta q_{2} - A\delta q_{1} - B\delta p_{1} + 2D_{3}a_{2s}\delta a_{2}^{\dagger} + 2D_{3}a_{2s}^{*}\delta a_{2} + \sqrt{2\gamma_{m2}}\delta p_{in}^2-\gamma_{m2}\delta p_{2}
\end{equation}

\begin{equation}
\dot{\delta a_{1}}= - i \omega_{1}\delta a_{1} + i D_{1}(q_{1s}\delta a_{1} + a_{1s}\delta q_{1}) - i D_{2}(a_{1s}\delta p_{1} + p_{1s}\delta a_{1}) - \kappa_{1}\delta a_{1} + \sqrt{2\kappa_{1}}\delta C_{in}^1
\end{equation}

\begin{equation}
\dot{\delta a_{2}}= -i  \omega_{2}\delta a_{2} + i D_{3}(q_{2s}\delta a_{2} + a_{2s}\delta q_{2}) - i D_{4}(a_{2s}\delta p_{2} + p_{2s}\delta a_{2}) - \kappa_{2}\delta a_{2} + \sqrt{2\kappa_{2}}\delta C_{in}^2
\end{equation}

On fourier transforming all operators and noise sources of equation 9 to 14 and solving in the frequency domain, the position fluctuations $\delta q_{1}(\omega)$ of the movable mirror M2 is obtained as -

\begin{eqnarray}
\delta q_{1}(\omega) &=& \frac{1}{d(\omega)}\{X(\omega)(\frac{a_{1s}^*}{\kappa_1- i (\omega - \Delta_1)}\delta C_{in}^1 + \frac{a_{1s}}{\kappa_1-i (\omega + \Delta_1)}\delta C_{in}^{1\dagger})2\sqrt{2\kappa_1} \nonumber \\
&+& Y(\omega)(\frac{a_{2s}^*}{\kappa_2- i (\omega - \Delta_2)}\delta C_{in}^2 + \frac{a_{2s}}{\kappa_2- i (\omega + \Delta_2)}\delta C_{in}^{1\dagger})2\sqrt{2\kappa_2} \nonumber \\
&+& (\sqrt{2 \gamma_{m1}}Z(\omega)\delta p_{in}^1 + \sqrt{2 \gamma_{m2}}T(\omega) \delta p_{in}^2)\}
\end{eqnarray}

 All the variables are defined in the appendix. In equation 14, the first two terms corresponding to $X(\omega)$ and $Y(\omega)$ gives rise the effect of radiation pressure whereas last two terms corresponding to $Z(\omega)$ and $T(\omega)$ originate because of the thermal noise. The coupling to the mirror shifts the cavity resonance frequency and changes the field inside the cavity in a way to induce a new stationary intensity. The shift in the cavity resonance is seen in the renormalized detunings $\Delta_{1}$ and $\Delta_{2}$. The change occurs after a transient time depending on the response of the cavity and the strength of the coupling to the mirrors. Now the spectrum of fluctuation of mirror can be defined as -

\begin{equation}
S_{q}(\omega) = \frac{1}{4\pi}\int d\Omega e^{- i(\omega+\Omega)t}<\delta q(\omega)\delta q(\Omega) + \delta q(\Omega)\delta q(\omega)>
\end{equation}

\begin{figure}[t]

\begin{tabular}{cc}
×\includegraphics{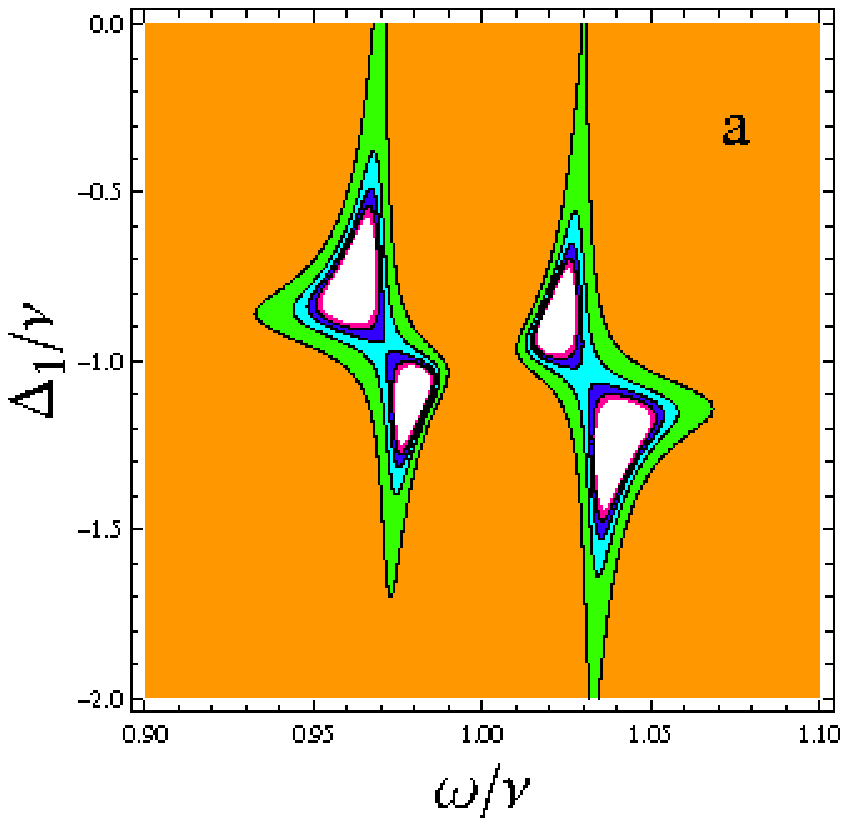}& \includegraphics{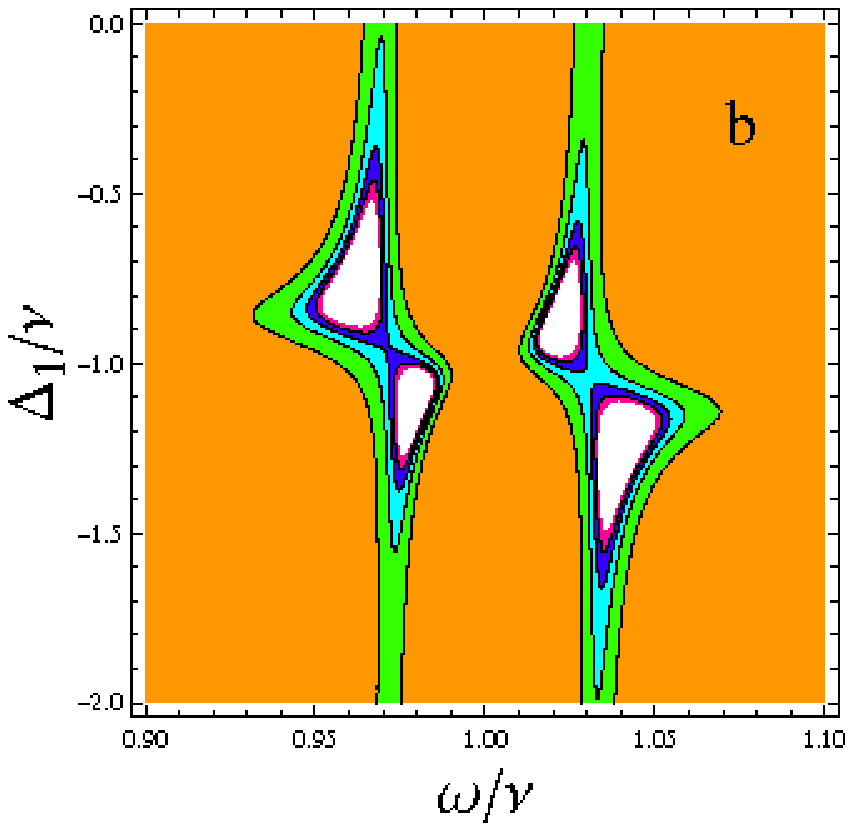}\\
 \includegraphics{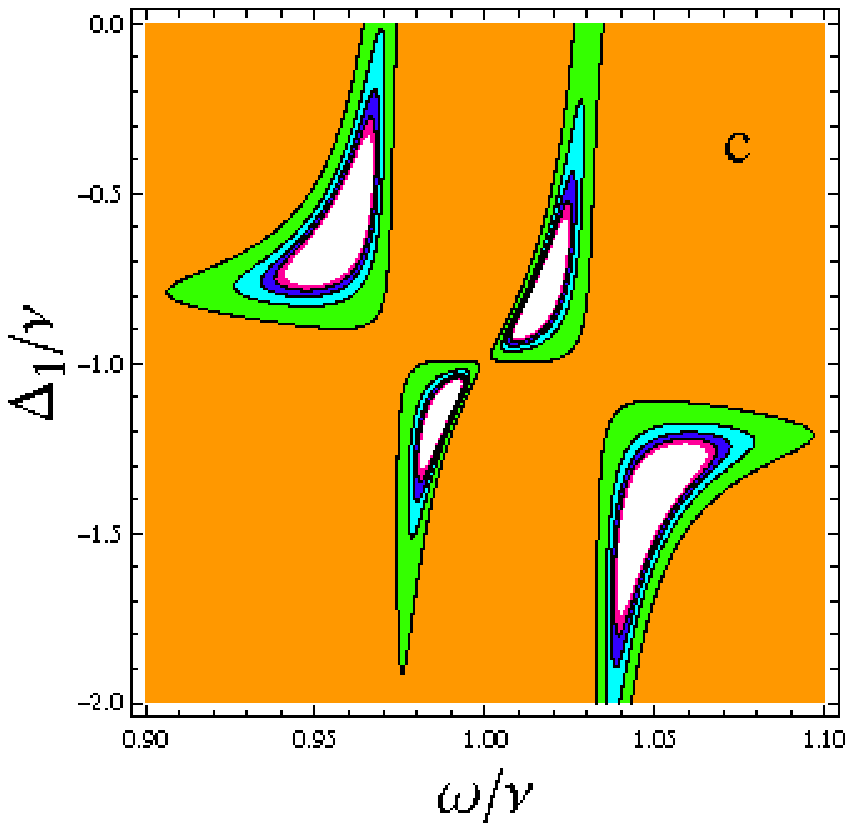} &\includegraphics{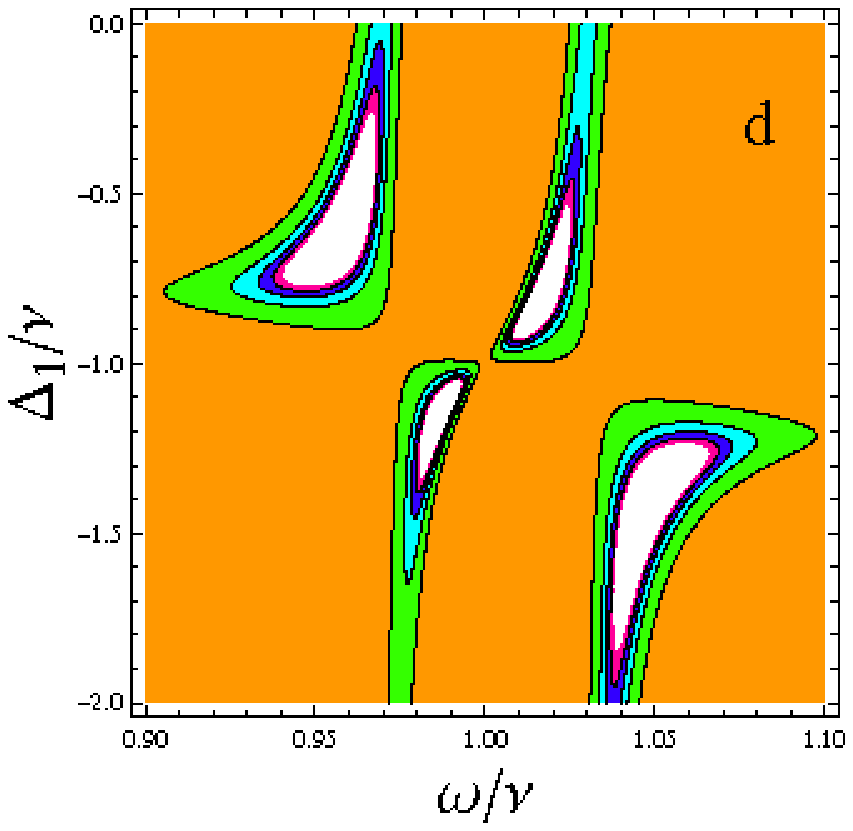}
\end{tabular}

\caption{Contour plot of the displacement spectrum $S_{q1}$ as a function of normalized effective detuning $\Delta_{1}$ for the following parameters: $\theta_1 = 0$, $\theta_2 =  \pi/2$,$\Delta_{2}/\nu=0$, $\gamma_1/\nu$ $=$ $\gamma_2/\nu$ $=$ $0.01$, $g_1/\nu$ $=$ $g_2/\nu$ $=$ $0.2$, $\nu_{12}/\nu$ $=$ $0.03$, $\kappa_1/\nu$ $=$ $\kappa_2/\nu$ $=$ $0.1$. (a): $|a_{1s}|^{2}=0.1$ and $|a_{2s}|^{2}=0.1$, (b):$|a_{1s}|^{2}=0.1$ and $|a_{2s}|^{2}=0.25$, (c): $|a_{1s}|^{2}=0.25$ and $|a_{2s}|^{2}=0.1$, (d):$|a_{1s}|^{2}=0.25$ and $|a_{2s}|^{2}=0.25$. The values of the color scheme is shown in Fig.5 in the appendix. }

\label{figure2}
\end{figure}

\begin{figure}[t]

\begin{tabular}{cc}
\includegraphics [scale=0.7] {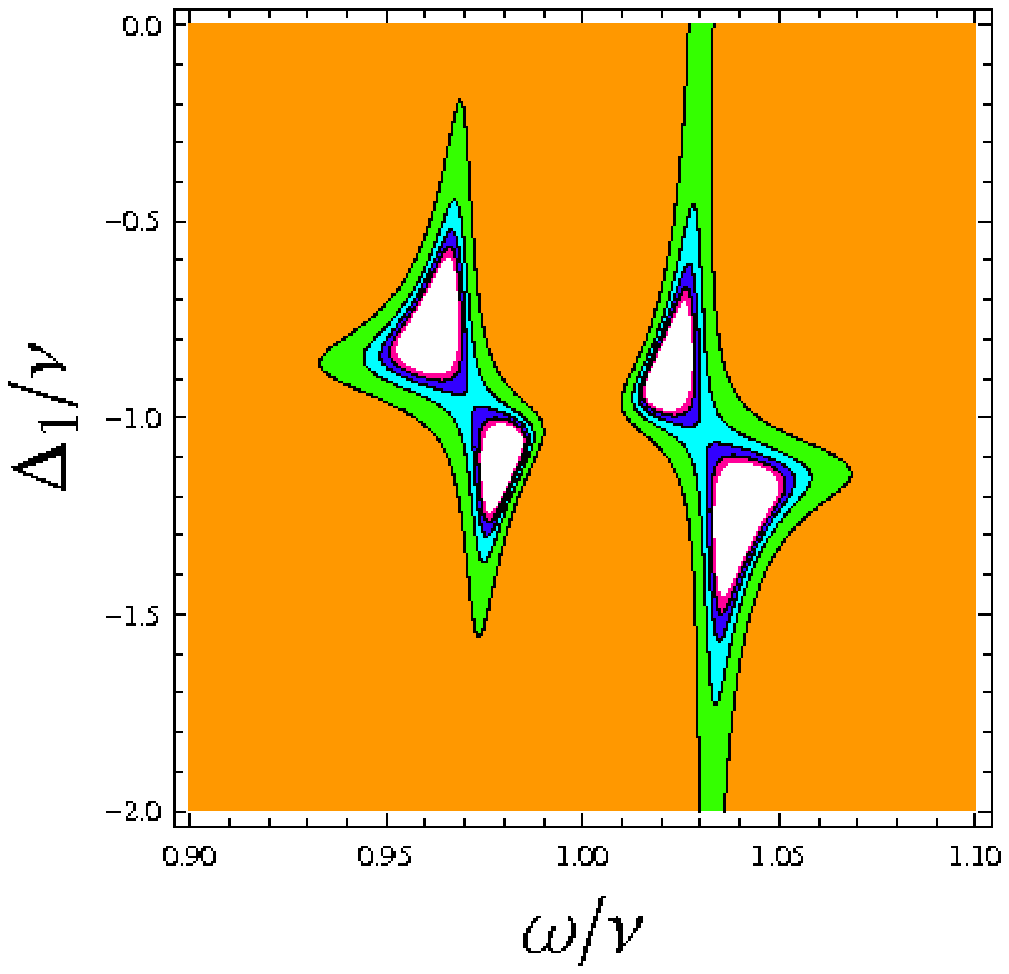}&\includegraphics [scale=0.7] {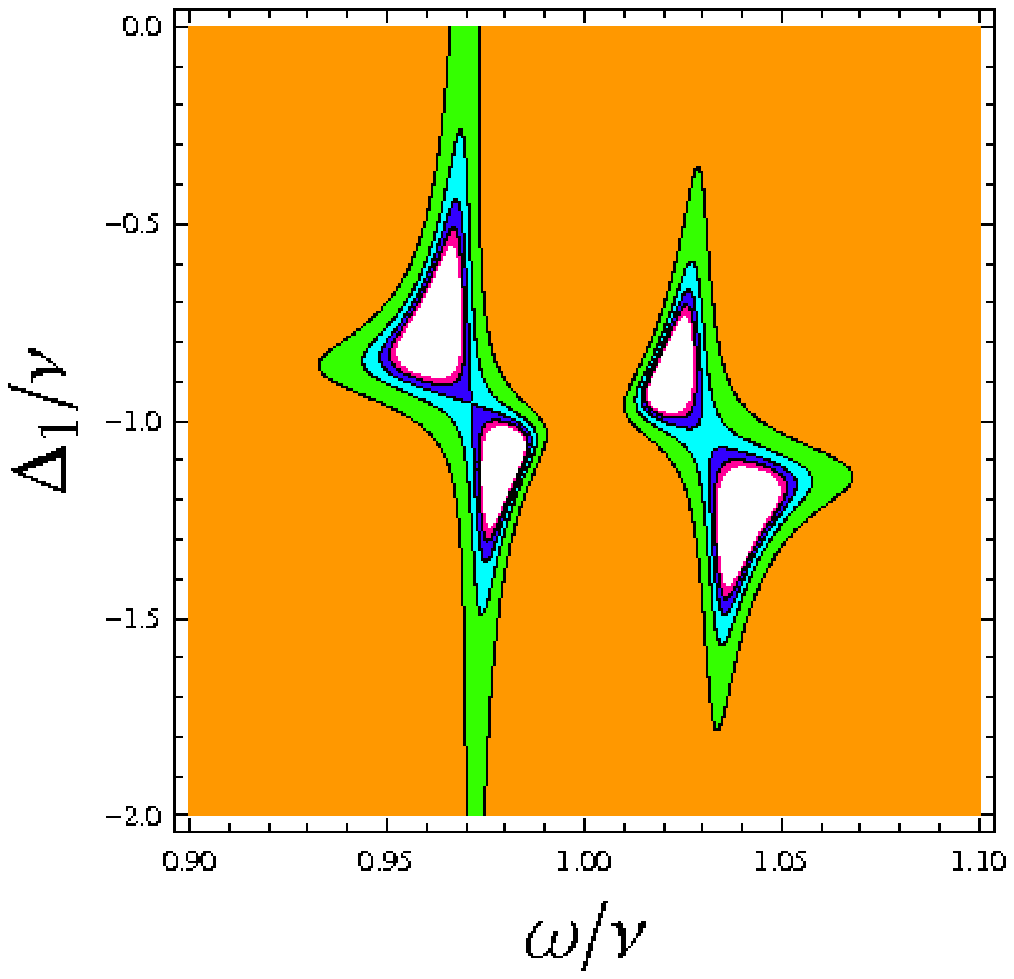}\\
\end{tabular}

\caption{ Contour plot of the displacement spectrum $S_{q1}$ as a function of dimensionless effective detuning $\Delta_{1}/\nu$ for the following parameters: $\Delta_{2}/\nu=0$, $\gamma_1/\nu$ $=$ $\gamma_2/\nu$ $=$ $0.01$, $g_1/\nu$ $=$ $g_2/\nu$ $=$ $0.2$, $\nu_{12}/\nu$ $=$ $0.03$, $\kappa_1/\nu$ $=$ $\kappa_2/\nu$ $=$ $0.1$, $|a_{1s}|^{2}=0.1$, $|a_{2s}|^{2}=0.1$,  $\theta_1 = 0$, $\theta_2 =  \pi/4$ (left plot), $\theta_1 = 0$, $\theta_2 = 3\pi/4$ (right plot).}

\label{figure3}
\end{figure}

The displacement spectrum of mirror M2 i.e, $S_{q1}(\omega)$ is finally obtained as -

\begin{eqnarray}
S_{q1}(\omega) &=& \frac{1}{d(\omega)d(-\omega)}[(\frac{4|a_{1s}|^2\kappa_1}{\kappa_{1}^2+(\omega - \Delta_1)^2}+ \frac{4|a_{1s}|^2\kappa_1}{\kappa_{1}^2+(\omega +\Delta_1)^2})X(\omega)X(-\omega) + (\frac{4|a_{2s}|^2\kappa_2}{\kappa_{2}^2+(\omega -\Delta_2)^2}  \nonumber \\
&+& \frac{4|a_{2s}|^2\kappa_2}{\kappa_{2}^2+(\omega +\Delta_2)^2})Y(\omega)Y(-\omega)+ (1+2n_T)(\gamma_{m1}^{2}Z(\omega)Z(-\omega) + \gamma_{m2}^{2}T(\omega)T(-\omega))];
\end{eqnarray}

Here we have used the commutation relation for $\xi$. In equation 16, the first two term due to radiation pressure contribution of the optical modes in the two cavities, whereas the last term is due to the thermal noise contribution from the two cavities.

In Fig. 2, we show the contour plot of the displacement spectrum $S_{q1}$ as a function of dimensionless effective detuning $\Delta_{1}/\nu$ for $\theta_1 = 0$, $\theta_2 =  \pi/2$,$\Delta_{2}/\nu=0$, $\gamma_1/\nu$ $=$ $\gamma_2/\nu$ $=$ $0.01$, $g_1/\nu$ $=$ $g_2/\nu$ $=$ $0.6$, $\nu_{12}/\nu$ $=$ $0.3$, $\kappa_1/\nu$ $=$ $\kappa_2/\nu$ $=$ $0.1$ for different values of the photon numbers in the two cavities, (a): $|a_{1s}|^{2}=0.1$ and $|a_{2s}|^{2}=0.1$, (b):$|a_{1s}|^{2}=0.1$ and $|a_{2s}|^{2}=0.25$, (c): $|a_{1s}|^{2}=0.25$ and $|a_{2s}|^{2}=0.1$, (d):$|a_{1s}|^{2}=0.25$ and $|a_{2s}|^{2}=0.25$. The values of the color scheme is shown in Fig.5 in the appendix. Clearly, four modes are visible corresponding to the two mechanical and two optical modes. The coupling of the cavity field fluctuations and the mirror fluctuations leads to splitting of the normal mode of each cavity into two modes (NMS). The NMS is associated with the mixing between the fluctuation of the cavity field around the steady state and the fluctuations of the mirror mode around the mean field. The origin of the fluctuations of the cavity field is the beat of the pump photons with the photons scattered from the mirrors. We observe from the displacement spectra that NMS is observed only in plots (a) and (b) where the photon number in first cavity is $|a_{1s}|^{2}=0.1$. Increasing the photon number in the first cavity destroys the NMS. We also note by comparing plots (b) and (c) that decreasing the photon number in the second cavity to $|a_{2s}|^{2}=0.1$ does not restore the NMS. An important point to note is that in order to observe the NMS, the energy exchange between the modes should take place on a time scale faster than the decoherence of each mode.  The parameter regime in which NMS appears implies cooling. For other values of the system parameters, the observation of NMS is prevented by the onset of the parametric instability. Therefore, a presence of NMS cannot be decoupled from the associated cooling which we discuss next where we calculate the effective temperature.

Energy exchange between the modes of the two cavities depends on the two phases $\theta_{1}$ and $\theta_{2}$. In Fig.3 we show the contour plot of the displacement spectrum $S_{q1}(\omega)$ as a function of dimensionless effective detuning for $\theta_1 = 0$, $\theta_2 =  \pi/4$ (left plot) and $\theta_1 = 0$, $\theta_2 = 3\pi/4$ (right plot). Clearly we observe energy exchange between the modes as we go from left to the right plot. Such energy exchange also implies that we can selectively cool one mirror at the expense of the other. In general it is known from basic physics that energy exchange between two mechanical oscillators takes place only for the anti-symmetric mode i.e when each mechanical oscillator is initially displaced from its position in opposite direction. In our case, such energy exchange between the two cavities can be achieved by tuning the phases $\theta_{1}$ and $\theta_{2}$.

We now calculate the effective temperature of the mirror $M2$. In order to calculate the effective temperature, we need the spectrum of the momentum of the mirror $M2$ in fourier space. In a similar manner as above, we can calculate the momentum spectrum of the mirror $M2$, which is found as:

\begin{eqnarray}
S_{p1}(\omega) &=& \frac{1}{t_{15}(\omega)t_{15}(-\omega)}[(\frac{4|a_{1s}|^2\kappa_1}{\kappa_{1}^2+(\omega - \Delta_1)^2}+ \frac{4|a_{1s}|^2\kappa_1}{\kappa_{1}^2+(\omega +\Delta_1)^2})t_{16}(\omega)t_{16}(-\omega) + (\frac{4|a_{2s}|^2\kappa_2}{\kappa_{2}^2+(\omega -\Delta_2)^2}  \nonumber \\
&+& \frac{4|a_{2s}|^2\kappa_2}{\kappa_{2}^2+(\omega +\Delta_2)^2})t_{17}(\omega)t_{17}(-\omega)+ (1+2n_T)t_{18}(\omega)t_{18}(-\omega)(\gamma_{m1}^{2} + \gamma_{m2}^{2})];
\end{eqnarray}

For a driven system, effective temperature can be defined as \citep{huang}:

\begin{equation}
T_{eff}=\dfrac{< \delta p^{2}>+<\delta q^{2}>}{2},
\end{equation}

where the variances are calculated as,

\begin{equation}
<\delta q^{2}>=\dfrac{1}{2 \pi}\int_{-\infty}^{\infty} S_{q}(\omega) d\omega
\end{equation}

\begin{equation}
<\delta p^{2}>=\dfrac{1}{2 \pi}\int_{-\infty}^{\infty} S_{p}(\omega) d\omega
\end{equation}

The equation for the effective temperature is one of our key results which tells how the temperature of one mirror depends on the various system parameters. Note that in general $\delta q^{2}$ $\neq$ $<\delta p^{2}>$. This implies that one does not have energy equipartition. This means that the steady state of the system is not, strictly speaking , a thermal equilibrium state. However, in order to get to the quantum ground state, both variances have to tend to $1/2$ and therefore energy equipartition has to be satisfied in the optimal regime close to the ground state.

\begin{figure}[h]
\hspace{-0.0cm}
\includegraphics [scale=1.0]{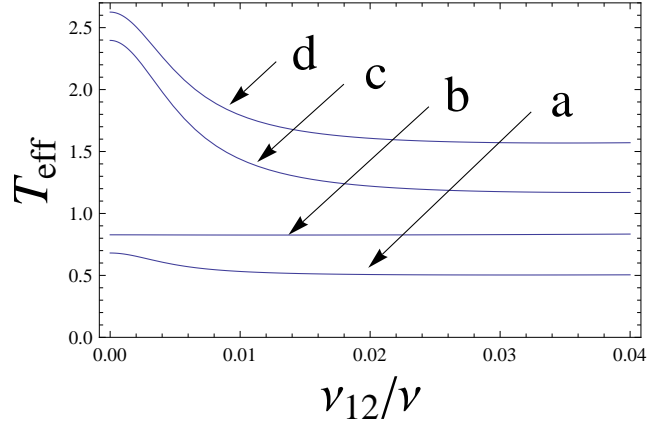}
\caption{ A plot of the effective temperature $T_{eff}$ of the mirror $M_{2}$ as a function of mirror-mirror coupling strength $\nu_{12}$ for four different values of $|a_{1s}|^{2}$ and $|a_{2s}|^{2}$. (a): $|a_{1s}|^{2}=0.1$ and $|a_{2s}|^{2}=0.1$, (b):$|a_{1s}|^{2}=0.1$ and $|a_{2s}|^{2}=0.25$, (c): $|a_{1s}|^{2}=0.25$ and $|a_{2s}|^{2}=0.1$, (d):$|a_{1s}|^{2}=0.25$ and $|a_{2s}|^{2}=0.25$. Other parameters are same as in Fig.2. }
\label{4}
\end{figure}

\begin{figure}[h]

\begin{tabular}{cc}
×\includegraphics{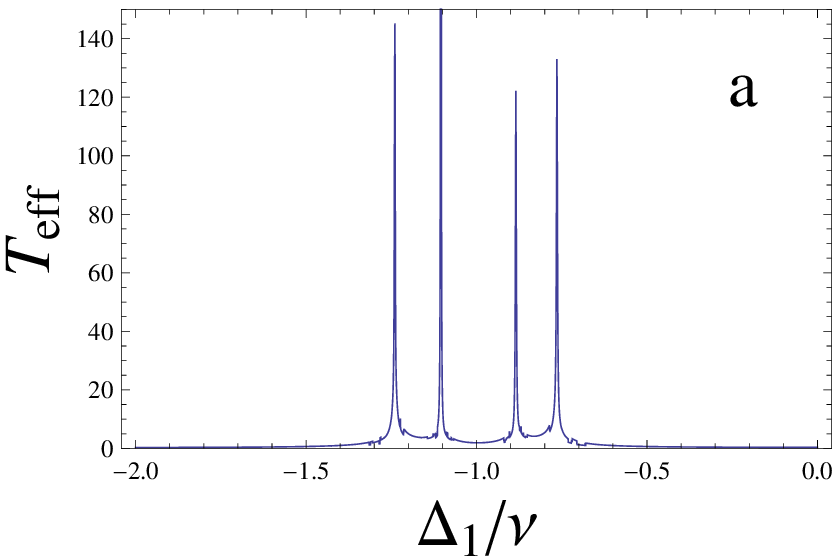}& \includegraphics{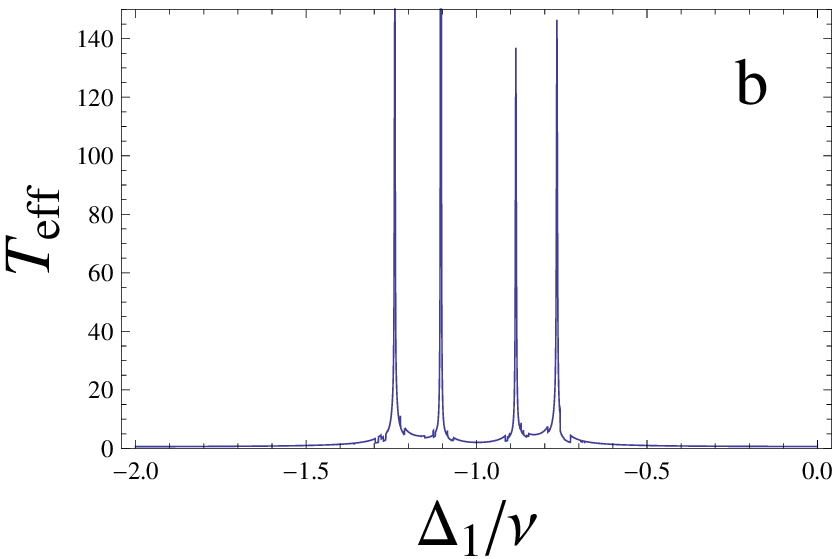}\\
 \includegraphics{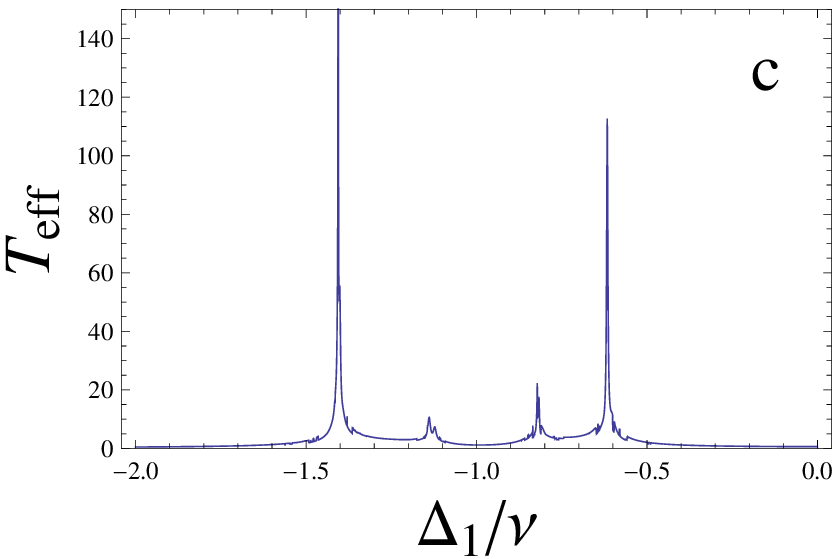} &\includegraphics{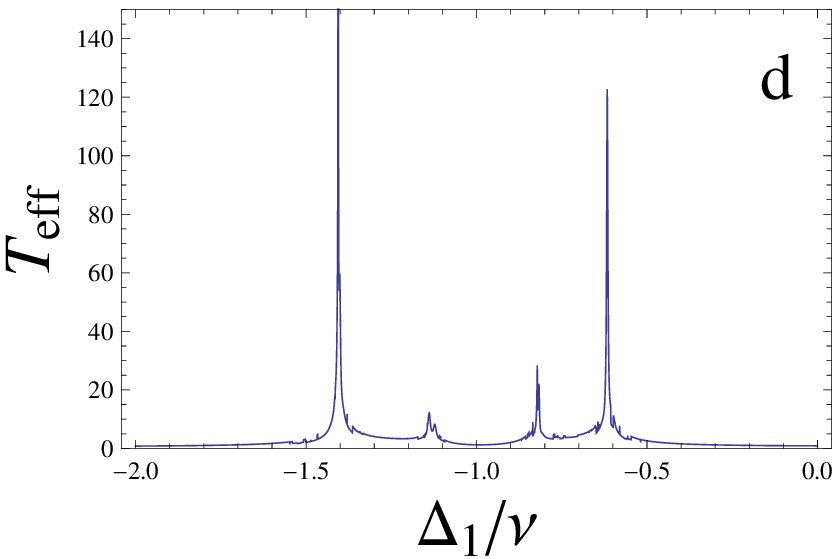}
\end{tabular}

\caption{ Plot of $T_{eff}$ of the mirror $M_{2}$ as a function of detuning $\Delta_{1}/\nu$ corresponding to the cases shown in Fig.2.Other parameters are same as in Fig.2. }

\label{figure5}
\end{figure}

 A plot of the effective temperature $T_{eff}$ of the mirror $M_{2}$ as a function of dimensionless mirror-mirror coupling strength $\nu_{12}/\nu$ for four different combinations of $|a_{1s}|^{2}$ and $|a_{2s}|^{2}$. (a): $|a_{1s}|^{2}=0.1$ and $|a_{2s}|^{2}=0.1$, (b):$|a_{1s}|^{2}=0.1$ and $|a_{2s}|^{2}=0.25$, (c): $|a_{1s}|^{2}=0.25$ and $|a_{2s}|^{2}=0.1$, (d):$|a_{1s}|^{2}=0.25$ and $|a_{2s}|^{2}=0.25$ is shown in Fig. 4. We clearly observe that minimum temperature is attained when both the cavities have low photon numbers. Increasing the photon number in any one of the cavity increases the temperature and the influence of $|a_{1s}|^{2}$ is more profound than $|a_{2s}|^{2}$. This is consistent with our earlier result on the NMS where we mentioned that the presence of the NMS also indicates cooling. Plots (a) and (b) in Fig.4 which show minimum temperature corresponds to plots (a) and (b) in Fig.2 which show NMS. Cooling of the mechanical mode of the mirror by the radiation pressure can be understood in thermodynamical sense. Radiation pressure couples the mirror to the optical cavity mode, which behaves as an effective additional reservoir for the mechanical oscillator. As a consequence, the effective temperature of the mirror mode  will be intermediate between the initial thermal reservoir temperature and that of the optical reservoir, which is practically zero due to the condition that the mean number of photons is extremely small. Therefore one can approach the mechanical ground state of the mirror when the number of photons is small. In our case, the mechanical mode of mirror $M_{2}$ not only couples to optical mode of first cavity but also to the optical mode of the second cavity via the mechanical mode of the mirror $M_{3}$. This explains why significant mechanical cooling of the mechanical mode is obtained when number of photons in both the cavities is low.
Fig.5 shows the $T_{eff}$ of the mirror $M_{2}$ as a function of detuning $\Delta_{1}/\nu$ corresponding to the cases shown in Fig.2. Interestingly we observe that the peaks in $T_{eff}$ corresponds exactly to the points in Fig.2 where the mirror displacement is maximum. These plots also illustrates energy exchange between the various modes as we vary mean photon numbers in the two cavities. In this system, the presence of the additional modes of the second cavity allows one to transfer energy from the mechanical mode of the mirror $M_{2}$ and the optical mode $a_{1}$ to the mechanical mode of the mirror $M_{3}$ and the optical mode $a_{2}$. This shows that both mechanical and optical cooling can be achieved by transferring energy from one cavity to the other. From the experimental point of view, the mirror's position can be measured by means of a phase-sensitive detection of the cavity output, which  is then fed back to the mirror by applying a force whose intensity is proportional to the time derivative of the output signal, and therefore to the mirror's velocity.

\section{Conclusion}
In this work, we have proposed a new technique to cool a harmonically oscillating mirror of an optical cavity mechanically coupled to another movable mirror of a second optical cavity. The system behaves as four coupled oscillators exchanging energy. Energy exchange can be coherently controlled by the phases of the opto-mechanical coupling strength. We find that a hybridization of the two oscillating mirrors with the fluctuations of the two driving optical fields occurs and leads to a splitting of the mechanical and optical fluctuation spectra. We also showed that normal mode splitting (NMS) leads to mechanical cooling. Significant mechanical cooling can be achieved by controlling the photon number in the two cavities. In addition, we demonstrate for the first time that by coupling two cavities, we can cool one cavity (both in the mechanical and optical sense) by transferring energy to the other. A continuous variable entanglement between the two mechanical modes could be used to improve the detection of weak classical forces in optomechanical devices as atomic force microscopes or gravitational wave detectors.

\section{Acknowledgements}

One of the authors Tarun Kumar thanks the University Grants Commission, New Delhi for the Junior Research Fellowship.

\section{Appendix}

\begin{figure}[h]
\hspace{-0.0cm}
\includegraphics [scale=1.0]{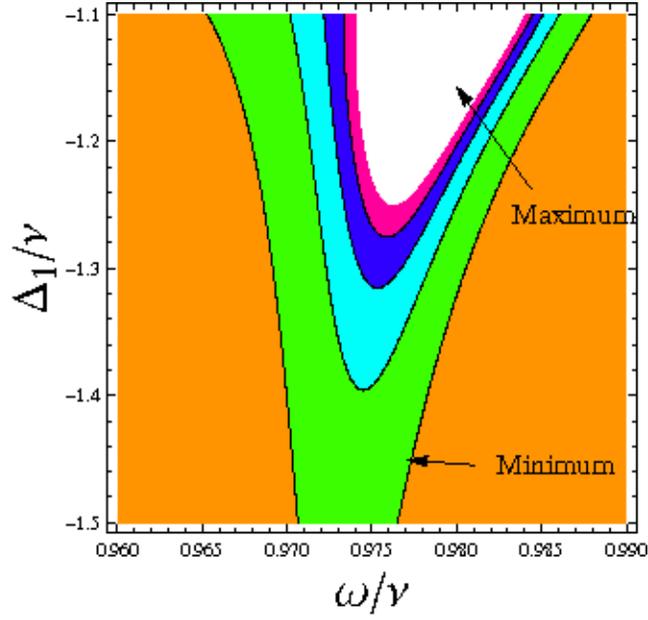}
\caption{Colour scheme for the contour plots }
\label{6}
\end{figure}

where, $d(w) = M'-\dfrac{R P}{K'}$;

 $X (\omega) = [\dfrac{R}{K'}(\dfrac{A}{C'}-\dfrac{B L}{F' C}) + (\dfrac{N}{C'}-\dfrac{A B}{F' C})] D_1 + D_2$;

 $Y (\omega)= [\dfrac{R}{K'}(\dfrac{A B}{C' F}+\dfrac{L}{F'}) + (\dfrac{A}{F'}+\dfrac{N B}{C' F})] D_3
+ D_4$;

$Z (\omega)= \dfrac{R}{K'}(\dfrac{A}{C'}-\dfrac{L B}{F' C}) + \dfrac{N}{C'}-\dfrac{A B}{F' C} $;

 $T (\omega) = \dfrac{R}{K'}(\dfrac{A B}{C' F}+\dfrac{L}{F'}) + \dfrac{A}{F'}+\dfrac{B N}{C' F} $;

$K' (\omega)=  K-\dfrac{A}{C'}(\dfrac{B G}{F}-A)-\dfrac{L}{F'}(G+\dfrac{A B}{C})$;

$C' (\omega)=  C+\dfrac{B^2}{F}; F' = F+\dfrac{B^2}{C}$;

$P (\omega)=  -B-\dfrac{L}{F'}(\dfrac{E B}{F'}+A)+\dfrac{A}{C'}(E-\dfrac{A B}{F})$;

$M' (\omega)=   M+\dfrac{A}{F'}(\dfrac{E B}{C}+A)-\dfrac{N}{C'}(E-\dfrac{A B}{F})$;

$R (\omega)= B+\dfrac{A}{F'}(\dfrac{A B}{C}+G)+\dfrac{N}{C'}(-A+\dfrac{B G}{F}) $;

$C (\omega)= \gamma_{m1}- i \omega + \dfrac{4 D_1D_2|a_{1s}|^2\Delta_1}{(\kappa_1- i \omega)^2+\Delta_{1}^2}$; $E (\omega) =-\nu_1+ \dfrac{4 D_{1}^2|a_{1s}|^2\Delta_1}{(\kappa_1- i \omega)^2+\Delta_{1}^2}$;

$F(\omega) = \gamma_{m2}- i  \omega + \dfrac{4 D_3D_4|a_{2s}|^2\Delta_2}{(\kappa_2- i \omega)^2+\Delta_{2}^2}$ ; $G (\omega) =-\nu_2+ \dfrac{4 D_{3}^2|a_{2s}|^2\Delta_2}{(\kappa_2- i \omega)^2+\Delta_{2}^2}$;

 $M (\omega)= - i \omega + \dfrac{4 D_1D_2|a_{1s}|^2\Delta_1}{(\kappa_1- i \omega)^2+\Delta_{1}^2}$ ; $N(\omega) =\nu_1+ \dfrac{4 D_{2}^2|a_{1s}|^2 \Delta_1}{(\kappa_1- i \omega)^2+\Delta_{1}^2}$;

$K (\omega)= - i \omega + \dfrac{4 D_3D_4|a_{2s}|^2\Delta_2}{(\kappa_2- i \omega)^2+\Delta_{2}^2}$ ; $L (\omega)=\nu_2+ \dfrac{4 D_{4}^2|a_{2s}|^2\Delta_2}{(\kappa_2- i \omega)^2+\Delta_{2}^2}$;

$\Delta_1 = \omega_1-q_{1s}D_1+p_{1s}D_2$ ; $\Delta_2 = \omega_2-q_{2s}D_3+p_{2s}D_4$;

$t_{15}(\omega)=t_{7}-\dfrac{t_{8}t_{12}}{t_{11}}$;

$t_{16}(\omega)=t_{9}+\dfrac{t_{8}t_{13}}{t_{11}}$;

$t_{17}(\omega)=t_{10}+\dfrac{t_{8}t_{14}}{t_{11}}$;

$t_{18}(\omega)=\dfrac{t_{8}}{t_{11}}$;

$t_{11}(\omega)=F-\dfrac{G t_{6}}{t_{4}}+\dfrac{A t_{3}}{t_{1}}$;

$t_{13}(\omega)=-\dfrac{D_{2}B G}{M t_{4}}-\dfrac{A D_{2}}{t_{1}}$;

$t_{12}(\omega)=\dfrac{G t_{5}}{t_{4}}-\dfrac{A t_{2}}{t_{1}}-B$;

$t_{14}(\omega)=D_{3}+\dfrac{D_{4} G}{t_{4}}-\dfrac{A D_{2}}{t_{1}}$;

$t_{7}(\omega)=C-\dfrac{E t_{2}}{t_{1}}+\dfrac{A t_{5}}{t_{4}}$;

$t_{8}(\omega)=B+\dfrac{E t_{2}}{t_{1}}-\dfrac{A t_{6}}{t_{4}}$;

$t_{9}(\omega)=D_{1}+\dfrac{D_{2} F}{t_{1}}+\dfrac{D_{2} A B}{M t_{4}}$;

$t_{10}(\omega)=\dfrac{D_{4} F B}{t_{1} K}-\dfrac{A D_{4}}{t_{4}}$;

$t_{1}(\omega)=M+\dfrac{B^{2}}{K}$;

$t_{2}(\omega)=N+ \dfrac{A B}{K}$;

$t_{3}(\omega)=A+ \dfrac{L B}{K}$;

$t_{4}(\omega)=K+\dfrac{B^{2}}{N}$;

$t_{5}(\omega)=A-\dfrac{B N}{M}$;

$t_{6}(\omega)=L-\dfrac{A B}{M}$

\end{document}